\documentclass[12pt]{article}
\usepackage{fullpage}
\usepackage{amsmath}
\usepackage{lineno}
\usepackage{url}
\usepackage{xcolor}

\usepackage{graphicx}

\title{\bf Intervention fatigue is the primary cause of strong secondary waves in the COVID-19 pandemic}

\author{\small Kristoffer Rypdal, Filippo Maria Bianchi, and  Martin Rypdal \\ {\small Department of Mathematics and Statistics} \\ {\small UiT -- The Arctic University of Norway}}
\date{}

\begin{document}
\maketitle

\begin{abstract}
As of November 2020, the number of COVID-19 cases is increasing rapidly in many countries. In Europe, the virus spread slowed considerably in the late spring due to strict lockdown, but a second wave of the pandemic grew throughout the fall. 
In this study, we first reconstruct the time evolution of the effective reproduction numbers ${\cal R}(t)$ for each country by integrating the equations of the classic Susceptible-Infectious-Recovered (SIR) model.
We cluster countries based on the estimated ${\cal R}(t)$ through a suitable time series dissimilarity.
The clustering result suggests that simple dynamical mechanisms determine how countries respond to changes in COVID-19 case counts. Inspired by these results, we extend the simple SIR model for disease spread to include a social response to explain the number $X(t)$ of new confirmed daily cases. 
As a first-order model, we assume that the social response is on the form $d_t{\cal R}=-\nu (X-X^*)$, where $X^*$ is a threshold for response. 
The response rate $\nu$ depends on whether $X^*$ is below or above this threshold, on three parameters $\nu_1,\;\nu_2,\,\nu_3,$, and on $t$.
When $X<X^*$, $\nu= \nu_1$, describes the effect of relaxed intervention when the incidence rate is low. When $X>X^*$, $\nu=\nu_2\exp{(-\nu_3t)}$, models the impact of interventions when incidence rate is high. The parameter $\nu_3$ represents the fatigue, \textit{i.e.}, the reduced effect of intervention as time passes. The proposed model reproduces typical evolving patterns of COVID-19 epidemic waves observed in many countries.
Estimating the parameters $\nu_1,\,\nu_2,\,\nu_3$ and initial conditions, such as ${\cal R}_0$, for different countries helps to identify important dynamics in their social responses.
One conclusion is that the leading cause of the strong second wave in Europe in the fall of 2020 was not the relaxation of interventions during the summer, but rather the general fatigue to interventions developing in the fall.
\end{abstract}

\section{Introduction}
The tendency of epidemics to return in repeated waves has been known since the 1918 Spanish flu \cite{Spanishflu}, and the recent COVID-19 pandemic is no exception. In November 2020, the history of reported daily cases, or incidence rate, of COVID-19 varies considerably across the world's regions. The broad picture is as follows \cite{ecdc}: The outbreak in China was practically over five weeks after a lockdown was imposed on January 23. Europe took over as the epicenter of the pandemic in late February. After lockdowns in most European countries in early March, followed by a gradual relaxation of these interventions, the first wave was over in the late spring. Incidence rates remained very low during the summer until they started to increase slowly in August. In late October, incidence rates higher than in March are common in Europe and are growing exponentially with a week's doubling time. The United States developed its first wave delayed by a week or two compared to Europe and a second and stronger wave throughout the summer. The country is now dealing with a third and even stronger wave. Many countries in South America, Africa, and South-East Asia are in the middle of (or have just finished) the first wave. On the other hand, a few countries, like New Zealand, Australia, Japan, and South Korea, have finished the second wave and managed to prevent it from becoming much stronger than the first one. Although there are a plethora of different wave patterns among the world's countries, one could hope that these patterns fall into a limited number of identifiable groups. 

In this paper, we do not claim that there is any epidemiological rule that states a pandemic evolving without social intervention must come in increasingly severe waves. On the contrary, the simple compartmental models devise the evolution of one single wave that finally declines due to herd immunity. We shall adopt the simplest of all such models here, the Susceptible-Infectious-Recovered (SIR) model, to describe the evolution of the epidemic state variables. However, in the SIR model, the effective reproduction number ${\cal R}$, the average number of new infections caused by one infected individual, is proportional to the fraction of susceptible individuals $S$ in the population. If initially ${\cal R}={\cal R}_0>1$, the daily number of new infections (incidence rate) will increase with time $t$ until $S$ has been reduced to the point where ${\cal R}$ goes below 1, and then decay to zero as $t\rightarrow \infty$.

At present, herd immunity is not an essential mechanism in the COVID-19 pandemic because the fraction of susceptible individuals is still close to 1 in most populations. Consequently, the time variation of the reproduction number is predominantly caused by changes in social behavior. Changes in virus contagiousness could also play a r\^ole, but we have not taken virus mutations into account in this paper. Thus, by adopting the approximation $S=1$ in the definition of the reproduction number, the SIR model reduces to a set of two first-order ordinary differential equations for the cumulative number of infected cases $J(t)$ and the instantaneous number of infectious individuals $I(t)$. These are the ``state variables'' of the epidemic, which are driven by the reproduction number ${\cal R}(t)$. The evolution of the epidemic state $[J(t),I(t)]$ can be computed as a solution to these equations if ${\cal R}(t)$ is known.

This paper's philosophy is to make the simplifying assumption that ${\cal R}(t)$ responds to the epidemic state. More precisely, that the rate of change of ${\cal R}(t)$ is a function of the rate of change of $J(t)$ depending on a set of parameters with distinct and straightforward interpretations that characterize the response. This mathematical relationship turns the SIR model into a closed model for the epidemic evolution, 
%which depends on these parameters. 
which depends on three parameters: i) the relaxation rate when incidence rate is low, ii) the intervention rate when incidence is high, and iii) a fatigue rate that gradually weakens the effect of interventions over time.
These parameters can be fitted to the incidence rate time series $X(t)=d_tJ(t)$ reported by different countries. 
The analysis of the fitted values of these parameters, allows to identify groups of countries with similar evolution of the epidemics and help to understand the most effective mechanisms controlling the epidemic's spread. 
One of our findings is stated in the paper's title; intervention fatigue is the primary mechanism that gives rise to the strong secondary waves emerging in many countries.\\

\noindent \textit{Paper outline.}\\
Section~\ref{sec:reserse_SIR} presents a method for reconstructing the ${\cal R}(t)$-profile from the observed time series for the daily incidence rate using a simple inversion of the SIR model.
The method's effectiveness is illustrated in Section~\ref {:reconstruction_examples} by application to selected representative countries. 
In Section~\ref{sec:clustering} we compute a dissimilarity measure between the reconstructed ${\cal R}(t)$-profiles for each country in the world.
Based on such a dissimilarity, we generate a dendrogram that hierarchically partitions countries according to their evolutionary paths of the epidemic.
Finally, in Section~\ref{sec:closed_model} we construct a self-consistent, closed model for the simultaneous evolution of $J(t)$ and ${\cal R}(t)$ and describe how this model can be fitted to the observed data for $d_tJ$ for individual countries.

In Section \ref{sec:clustering_results} we synthesize the reconstructed ${\cal R}$-curves for the majority of the world's countries and use the dendrogram to group them into seven clusters, which are also shown on a World map. 
The features characterizing each cluster are analyzed and discussed. 
Section~\ref{sec:params_exploration} illustrates the scenarios of the epidemic evolution that can be derived by solving the equations of the proposed closed model for different sets of parameters. 
The proposed model's effectiveness is empirically validated in Section~\ref{sec:fit}, where the model parameters are numerically fit to the incidence data for some selected countries exhibiting different characteristic patterns of epidemic evolution.

The possible implications of these results for COVID-19 strategic preparedness and response plans are discussed in Section \ref{sec:discuss}.

%%%%%%%%%%%%%%%%%%%%%%%%%%%%%%%%%%%%%%%%%%%%%%%%%%%%%%%%%%%%%%%%%%%%%%%%%%%%%%%%%%%%%%%%%%%%%%%%%%%%%%%%%%%%%%%%%%%%%%%
%%% ESTIMATION OF R FROM DATA
%%%%%%%%%%%%%%%%%%%%%%%%%%%%%%%%%%%%%%%%%%%%%%%%%%%%%%%%%%%%%%%%%%%%%%%%%%%%%%%%%%%%%%%%%%%%%%%%%%%%%%%%%%%%%%%%%%%%%%%
\section{Methods}
\label{sec:Methods}

\subsection{Estimating the reproduction number from incidence rate data}
\label{sec:reserse_SIR}
Let $S$ be the fraction of susceptible individuals in a population, $I$ the fraction of infectious, and $R$ the fraction of individuals ``removed" from the susceptible population (\textit{e.g.}, recovered, isolated, or deceased individuals). A simple model describing the evolution of these variables is the classical SIR-model \cite{SIR}, 
\begin{linenomath*}
\begin{eqnarray} 
\label{1}
\frac{dS}{dt}&=&-\beta IS, \\ 
\label{2}
\frac{dI}{dt}&=&\beta IS-\alpha I, \\ 
\label{3}
\frac{dR}{dt}&=&\alpha I, 
\end{eqnarray}
\end{linenomath*}
where $\alpha$ is the rate by which the infected are isolated from the susceptible population. Another interpretation of $\alpha$ is that $\alpha^{-1}$ is the average duration of the period an individual is infectious, which essentially depends only on the properties of the pathogen. As long as these do not change significantly, $\alpha$ will remain constant in time. In this paper we use $\alpha=1/(8 \text{ days})$ but our results are not sensitive to this choice. The coefficient $\beta$, on the other hand, is the rate by which the infection is being transmitted. It evolves in time as societal interventions change. It is also influenced by behavioral changes in the susceptible population, such as eliminating superspreaders. 
The effective reproduction number is defined as 
\begin{linenomath*}
\begin{equation}
{\cal R}(t) \equiv \frac{\beta(t)}{\alpha}\label{4}
\end{equation}
\end{linenomath*}
and can be interpreted as the average number of new infections caused by an infected individual over the infectious period $\alpha^{-1}$.

The coupled system given by Eqs.\;(\ref{1}) and (\ref{2}),  with initial conditions $S_0$ and $I_0$, constitutes a closed nonlinear initial value problem.  Eq.\;(\ref{3}) is not a part of this system since it is trivially integrated to yield the removed population $R(t)$ once $I(t)$ is known.

The method developed in this paper is valid for an infectious disease with a new pathogen which is transmitted by  contact between infectious and susceptible individuals. This implies that there is practically no immunity in the population from the  start of the epidemic and we shall assume that this herd immunity is low throughout the period for which we estimate the reproduction number. In other words, we shall assume that $S\approx 1$, and hence that the cumulative fraction $J=1-S$ of infected individuals is always much less than unity ($J\ll 1$). By introducing $S=1-J$ in Eqs.\;(\ref{1}) and (\ref{2}), and by neglecting the term $\beta IJ$ compared to the term $\beta I$ in Eq.\;(\ref{1}), these two equations reduce to a linear model for $J$ and $I$,
\begin{linenomath*}
\begin{eqnarray} \label{5} 
\frac{dJ}{dt}&=&\alpha {\cal R}(t) I \\ 
\label{6} 
\frac{dI}{dt}&=&\alpha[{\cal R}(t) -1] I\,. 
\end{eqnarray}
\end{linenomath*}
Note that $\gamma_I(t)=I^{-1}dI/dt=\alpha [{\cal R}(t)-1]$ is the relative growth rate for the instantaneous number of infectious individuals $I(t)$, which is positive when ${\cal R}(t)>1$ and negative when ${\cal R}(t)<1$. 
By integrating Eq.\;(\ref{6}) and by inserting the result on the right hand side of Eq.\;(\ref{5}), we obtain that the daily number of new infections $dJ/dt$ is determined by the initial  $I_0$ and  the  history of ${\cal R}(t)$ on the interval $(0,t)$; 
\begin{linenomath*}
\begin{equation}
 \frac{dJ}{dt}=\alpha I_0{\cal R}(t)\exp{\Bigg{(}\int_0^t\alpha[{\cal R}(t')-1)] dt'\Bigg{)}}. \label{7} 
\end{equation}
\end{linenomath*}
What we are interested in here, however, is the inverse relationship; suppose the evolution of $dJ/dt$ is known, how do we find the evolution of the reproduction number ${\cal R}(t)$? 

By using Eq.\;(\ref{5}) to replace $\alpha {\cal R}I$ by $d_tJ$ in Eq.\;(\ref{6}), the latter can be integrated to yield,
\begin{linenomath*}
\begin{equation}
    I(t)=I_0 e^{-\alpha t}+\int_0^t e^{-\alpha (t-t')}d_{t'}J\; dt', \label{8}
\end{equation}
\end{linenomath*}
which allows us to compute ${\cal R}(t)$ from Eq.\,(\ref{5});
\begin{linenomath*}
\begin{equation}
    {\cal R}(t) = \frac{d_tJ}{\alpha I} = \frac{1}{\alpha}\frac{d_tJ}{ I_0 e^{-\alpha t}+\int_0^t e^{\alpha (t'- t)}d_{t'}J\; dt'}.\label{9}
\end{equation}
\end{linenomath*}

Provided a time series for $J(t)$ is available, we can approximate $d_tJ$ as a finite difference and the integral in Eq.\;(\ref{9}) as a discrete sum. 
This sum gives us a fast and direct algorithm to estimate ${\cal R}(t)$.

\subsection{{\cal R}(t)-reconstructions for individual countries}
\label{sec:reconstruction_examples}
Since we do not have actual measurements of the cumulative number of infected, to estimate the $\mathcal{R}(t)$-curves for each country using Eq.\;(\ref{9}) we rely on the number of confirmed cases as a proxy for $d_tJ(t)$. 
Specifically, we assume that the incidence rate $d_tJ(t)$ is proportional to the daily number of confirmed cases $X(t)$. 

\begin{figure}[t!]
	\centering
\includegraphics[width=16cm]{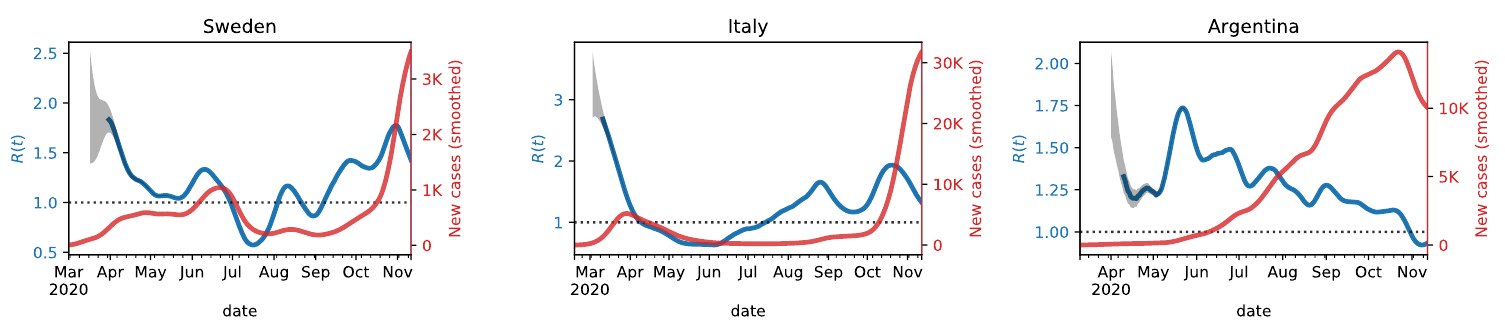}
    \caption{\footnotesize $\mathcal{R}(t)$ estimated with the proposed method from the time series of new daily cases reported in Sweden, Italy, and Argentina. Red curves are reported incidence rates, and the blue curves the reconstructed reproduction numbers.}
	\label{fig:examples}
\end{figure}

The time series $X(t)$ of new daily cases reported for each country are taken from the Our World in Data database\footnote{\url{https://ourworldindata.org/coronavirus-source-data}}.
Figure~\ref{fig:examples} shows three examples of $\mathcal{R}(t)$ estimated using Eq.\;(\ref{9}) from the new daily cases $X(t)$ reported by Sweden, Italy, and Argentina. 
The initial values assumed by $\mathcal{R}(t)$ are affected by different choices of $I_0$. The transient effect given by the initial conditions quickly vanishes as $t$ increases, and $\mathcal{R}(t)$ converges to a stable solution since the first term in the denominator of Eq.\;(\ref{9}) goes exponentially to zero. To compute the results, we generated several initial conditions for $I_0$ in a reasonable range, and we discarded the transient phase, depicted as a gray area in Fig.~\ref{fig:examples}. In these plots we have taken into consideration the delay between the date of infection and the reported positive tests, which may amount to approximately one week. Thus, the actual ${\cal R}(t)$ curves should be shifted towards the left by approximately this amount.

The three countries in Fig.~\ref{fig:examples} are characterized by a different evolution of the epidemics.
Sweden had a long first wave that peaked in the middle of June and the second wave started when the first one was not completely over.
This is reflected in the estimated $\mathcal{R}(t)$, which stays for a long time interval above 1.
Italy had a first wave stronger than other countries, that was brought down completely due to the lockdown.
The estimated $\mathcal{R}(t)$ starts from very high values and quickly goes below 1 by the beginning of April.
Finally, the number of new cases in Argentina kept growing very slowly, but consistently, until the middle of October and there are no two distinct waves like in many other countries. 
Consequently, the $\mathcal{R}(t)$ is characterized by values that are slightly above 1 until the beginning of November.

\subsection{Cluster analysis of {\cal R}(t)-curves}
\label{sec:clustering}

Rather than presenting reconstructions of ${\cal R}(t)$  case by case for all of the world's countries, it would bring more insight to combine them in groups of ${\cal R}(t)$-curves according to some common features and then analyze the characteristics of each group.
Therefore, we follow such an indirect approach where we first cluster the $\mathcal{R}(t)$ curves of different countries and then analyze the clustering partition and the representatives of each cluster.
The cornerstone of each clustering algorithm is the computation of a dissimilarity measure between the data samples.
Since we are dealing with with sequential data, we leverage on a dissimilarity measure $\delta_{i,j} = d(x_i, x_j)$ that yields a real number $\delta_{i,j}$ proportional to the discrepancy between the time series $x_i$ and $x_j$.

A large variety of time series dissimilarity measures have been proposed in the literature, including those based on statistical methods~\cite{de2011tail}, signal processing~\cite{chan1999efficient}, kernel methods~\cite{mikalsen2018time}, and reservoir computing~\cite{bianchi2020reservoir}.
In this paper, we adopt the Dynamic Time Warping (DTW) distance~\cite{keogh2005exact}, which is an efficient and well-known algorithm that computes the dissimilarity between two sequences as the cost required to obtain an optimal match between them. The cost is computed as the sum of absolute differences between a set of indices in the two time series.
DTW allows similar shapes to match, even if they are out of phase or, in general, not perfectly synchronized along the time axis.

From the dissimilarity $\delta_{i,j}$ between countries $i$ and $j$, it is possible to compute a clustering partition, where similar $\mathcal{R}$-time-series are assigned to the same cluster.
Several approaches can be used to generate the clusters~\cite{aghabozorgi2015time}. 
We opted for a hierarchical clustering method~\cite{cohen2018hierarchical}, which gradually joins data samples together by increasing the maximum radius of the clusters' $\delta_\text{max}$.
One of the main advantages of hierarchical clustering is the possibility of generating a dendrogram, which allows to visually explore the structure of the clustering partition at different resolution levels.

%%%%%%%%%%%%%%%%%%%%%%%%%%%%%%%%%%%%%%%%%%%%%%%%%%%%%%%%%%%%%%%%%%%%%%%%%%%%%%%%%%%%%%%%%%%%%%%%%%%%%%%%%%%%%%%%%%%%%%%
%%% CLOSED EPIDEMIC MODEL
%%%%%%%%%%%%%%%%%%%%%%%%%%%%%%%%%%%%%%%%%%%%%%%%%%%%%%%%%%%%%%%%%%%%%%%%%%%%%%%%%%%%%%%%%%%%%%%%%%%%%%%%%%%%%%%%%%%%%%%
\subsection{A closed model for model for the  epidemic evolution}
\label{sec:closed_model}
The SIR-model does not constitute a closed model for the evolution of $J(t)$, $I(t)$, and ${\cal R}(t)$. Eqs.\;(\ref{5}--\ref{6}) describe the dynamics of the epidemic state variables $J(t)$ and $I(t)$ when the evolution of the  social state represented by ${\cal R}(t)$ is given. Eq.\;(\ref{9}) is nothing but an inverse of this relationship and should not be interpreted as a social response of ${\cal R}(t)$ to changes in the epidemic state variables. A closed model can only be obtained by adding an equation describing such a response. 
While a simple dynamical model can not reflect the whole complexity of the social response, it may still provide some useful insight. 

We shall represent this response by assuming that the rate of change  $d_t{\cal R}(t)$ is a function of the incidence rate $X(t)=d_tJ(t)$, 
and that this function is positive when $X(t)$ is below a threshold $X^*$ and negative when it is above that threshold. When the incidence rate is low, society responds by relaxing restrictions, and the reproduction number increases. When the incidence rate exceeds the threshold $X^*$, restrictions are introduced that make $d_t{\cal R}(t)$ to change sign from positive to negative.

In the following, it is convenient to introduce a dimensionless time variable $t'= \alpha t$, which allows us to formulate the differential equations as functions of the mean infectious time $\alpha^{-1}$ (which becomes the new time unit) rather than days.  
Accordingly, Eq.\;(\ref{5}) can be written as
\begin{linenomath*}
\begin{equation}
 X(t')\equiv d_{t'} J= {\cal R} I,   \label{16}
\end{equation} 
\end{linenomath*}
and we have a closed model for $I(t')$ and ${\cal R}(t')$ in the form of the dynamical system,
\begin{linenomath*}
\begin{eqnarray}
\label{17} 
\frac{d{\cal R}}{dt'} &=& f(X)=f({\cal R} I),\\ 
\label{18}
\frac{dI}{dt'} &=& ({\cal R}-1)I, 
\end{eqnarray}
\end{linenomath*}
where $f(X)$ is assumed to be a differentiable function which is decreasing in a neighborhood of $X^*$ and with $f(X^*)=0$. The system has a fixed point in ${\cal R}=1$ and  $I = X^*$. In this state, the number of infected stays constant at the threshold value.

By linearization of $f(X)$ around the fixed point ${\cal R}=1$ and  $I = X^*$, and by introducing the rate constant $\nu = -(1/2)X^*f'(X^*)>0$, the system reduces to
\begin{linenomath*}
\begin{eqnarray}
\label{19}
\frac{d\Delta {\cal R}}{dt'} &=& -2\nu  (\Delta {\cal R} + \Delta \tilde{I}+ \Delta {\cal R}\Delta \tilde{I}) \\ 
\label{20}
\frac{d\Delta \tilde{I}}{dt'} &=&  \Delta {\cal R}(1+\Delta \tilde{I}), 
\end{eqnarray}
\end{linenomath*}
where we have introduced $\Delta {\cal R}=1-{\cal R}$ and the normalized number of infected $\tilde{I}=I/X^*=1+\Delta \tilde{I}$.  This nonlinear dynamical system has a stable fixed point in $(\Delta {\cal R},\Delta \tilde{I})=(0,0)$.

\subsubsection{The damped harmonic oscillator model}

 In this section we demonstrate that if $X$ is close to the threshold value $X^*$ and ${\cal R}$ is close to 1, the linearization of Eqs.\;(\ref{17}) and (\ref{18}) leads to the equation for a damped harmonic oscillator. The purpose is to show analytically under what circumstances a damped oscillation is a natural time-asymptotic state of the epidemic. In Section \ref{nonlinear model} we argue that the model needs to be generalized to yield realistic descriptions of epidemic curves in most countries and, hence, the present section may be skipped  without losing anything essential. 
 
 In the vicinity of the stable state $(\Delta {\cal R},\Delta \tilde{I})=(0,0)$, linearization yields the damped, harmonic oscillator equation, 
\begin{linenomath*}
\begin{equation}
\frac{d^2 \Delta \tilde{I}}{dt'^2}  + 2\nu  \frac{ d\Delta \tilde{I}}{dt'} + (\omega^2+\nu^2) \Delta \tilde{I} = 0,
\label{21}
\end{equation}
\end{linenomath*}
where $\omega^2  = 2\nu -\nu^2$.
For $\nu<2$, the general solution is the damped oscillator
\begin{linenomath*}
\begin{equation}
\Delta \tilde{I}(t') = A \, e^{-\mu t'} \cos(\omega t' + \varphi),    \label{22}
\end{equation}
\end{linenomath*}
where $A$ and $\varphi$ are integration constants and $\mu\equiv\nu$,
and for $\nu\geq 2$ the non-oscillatory strongly damped solution which for large $t'$ goes as
\begin{linenomath*}
\begin{equation}
  \Delta  \tilde{I}(t')=Be^{-\mu^{(-)} t'}+Ce^{-\mu^{(+)}t'}, \label{23}
\end{equation}
\end{linenomath*}
where $B$ and $C$ are constants of integration and $\mu^{(\pm)}\equiv \nu(1\pm \sqrt{1-2/\nu})$. From Eq.\;(\ref{20}),  we have 
\begin{linenomath*}
\begin{equation}
    \Delta {\cal R}=\frac{d\Delta \tilde{I}}{dt'}, \label{24}
\end{equation}
\end{linenomath*}
and from Eq.\;(\ref{16}),
\begin{linenomath*}
\begin{equation}
\Delta X(t')=\Delta(\tilde{I}(t'){\cal R}(t'))  \approx  \Delta {\cal R}(t') + \Delta \tilde{I}(t')=\frac{d\Delta \tilde{I}}{dt'}+\Delta \tilde{I}, \label{25}
\end{equation}
\end{linenomath*}
which means that $\Delta I$, $\Delta {\cal R}$, and $\Delta X$ experience the same damped oscillations with some phase shifts, or the same strongly damped solutions.

The frequency $\omega$ (for $0<\nu<2$) and the damping  rate $\mu$ (for $0<\nu<\infty$) are plotted against the parameter $\nu$ in Figure 1. The oscillation frequency and the damping rate are of comparable magnitude for $\nu<1$, but the damping dominates in the interval $1<\nu<2$. For $\nu>2$, the damping rate decreases towards 1 as $\nu$ increases.

\begin{figure}[t!]
\centering
\includegraphics[width=8 cm]{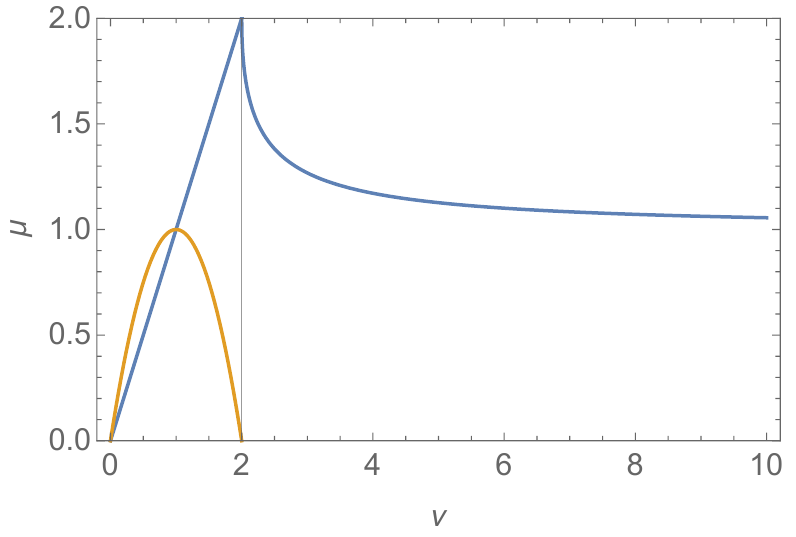}
\caption{The yellow curve shows the frequency $\omega=\sqrt{2\nu-\nu^2}$, and the blue curve the damping rate $\mu=\nu$ for $\nu<2$ and $\mu=\nu(1-\sqrt{1-2/\nu}-1)$ for $\nu\geq 2$.}
\end{figure}  

The most rapid control of the epidemic is obtained when $\mu\approx 2$, but we also have reasonably rapid control for larger $\nu$. Slower damping takes place when $\nu<1$, and slower the lower $\nu$. Hence, what really should be avoided is $\nu$ much less than 1.
The  linearization of  $f(X)$ around $X^*$ in Eq.\;(\ref{17}) yields
\begin{linenomath*}
\begin{equation}
    \frac{d{\cal R}}{dt'}=-\nu\Delta \tilde{X}, \label{26}
\end{equation}
\end{linenomath*}
where $\tilde{X}=X/X^*$ is the incidence rate normalized to its threshold value. which shows that $\nu$ is a measure of how fast the rate of change in ${\cal R}$ responds to the deviation of $\tilde{X}$ from its threshold value $\tilde{X}^*=1$. If $\nu\ll 1$ , then $d_t{\cal R}$ responds slowly to $\Delta \tilde{X}$, \textit{i.e.}, there is a slow social response to the rise or decay of the incidence. Eq.\;(\ref{21}) and Figure~\ref{fig:oscillator} then yields a damped oscillation with envelope that decays exponentially at a rate $\mu=\nu$. A characteristic duration of the epidemic is $\tau\equiv \nu^{-1}$ and the characteristic time scale of the oscillation  is $T =\omega ^{-1}=1/\sqrt{2\nu-\nu^2}$. Since the ratio between the two is $T/\tau =\sqrt{\nu/(2-\nu)}$, we observe that the oscillation scale $T$ is  longer than the decay time $\tau$ for all $\nu>1$. Hence, this model suggests that oscillatory behavior and a long duration of the epidemic are features we expect to observe when the social response is slow ($\nu<1$).

\subsubsection{A nonlinear, three-parameter oscillator model}
\label{nonlinear model}
Although the linear oscillator model gives some insight into the mechanism that makes the epidemic return in repeated waves, there is an obvious lack of realism. One is to neglect the terms containing the product $\Delta {\cal R}\Delta \tilde{I}$ in Eqs.\;(\ref{19})-(\ref{20}), since neither $\Delta {\cal R}$ nor $\Delta \tilde{I}$ are, in general, small. 
There is also little reason to expect that the rate of change $\nu$ is the same below and above the social response threshold. 
Below the threshold, ${\cal R}$ increases because of the intervention's termination and because the population relaxes. The more relaxed, the larger $\nu$, so let us denote this parameter $\nu_1$ the ``relaxation rate''. Above the threshold, ${\cal R}$ decreases because of the interventions aiming to strike the epidemic down. Stronger intervention translates into a larger ``intervention rate'' $\nu_2$. Even with these generalizations, the model will still give a damped, nonlinear oscillation and, hence, is unable to describe a situation where the second wave is stronger than the first. 
A generalization which may cover such a situation is to let the intervention rate decay with time, for instance, exponentially, such that we have an ultimate model for $\nu(t')$ on the form
\begin{linenomath*}
\begin{equation}
    \nu(t')=\nu_1\theta(-\Delta\tilde{X})+\nu_2e^{-\nu_3 t'}\theta(\Delta \tilde{X}), \label{27}
\end{equation}
\end{linenomath*}
where $\theta(x)$ is the unit step function. The parameter $\nu_3$ can be thought of as a ``fatigue rate'', \textit{i.e.}, the rate at which the strike-down rate is reduced because the population is becoming increasingly tired of interventions and restrictions.

Note also that the time dependence of the reproduction number in this model is independent of the response threshold $X^*$. This is because $X^*$ has been eliminated in  Eqs.\;(\ref{19})-(\ref{20}) through normalization of the variables. The un-normalized variables $I=X^*\tilde{I}$ and $X(t)$ are, of course, proportional to $X^*$ and emphasizes the importance of a low tolerance threshold for social intervention.

\subsubsection{Fitting model parameters to the observed incidence data}
To validate the effectiveness of the proposed model in describing real data, we fit the three parameters $\nu_1, \nu_2, \nu_3$ with a numerical optimization routine that minimizes the discrepancy between the time series of reported new daily cases $X(t)$ and those generated by the model.
We constrained $\nu_1 \in [0, \infty]$, while the other two parameters are unbounded.
Besides the three model parameters, we also optimize with a grid search the following hyperparameters:
the initial reproduction number $\mathcal{R}_0$ searched in the interval $[1.0, 3.0]$ and the value $X^{*} = X(t^{*})$ for each country with $t^{*}$ searched in the interval [15 January, 31 March].
As initial conditions for $\nu_1, \nu_2, \nu_3$ in the optimization routine, we used the values $[0.1, 0.1, 0.1]$.

\section{Results and discussion}
\label{sec:results}

\subsection{Results of the cluster analysis of {\cal R}(t)-curves}
\label{sec:clustering_results}

\begin{figure}[t!]
    \centering
 \includegraphics[width=14 cm]{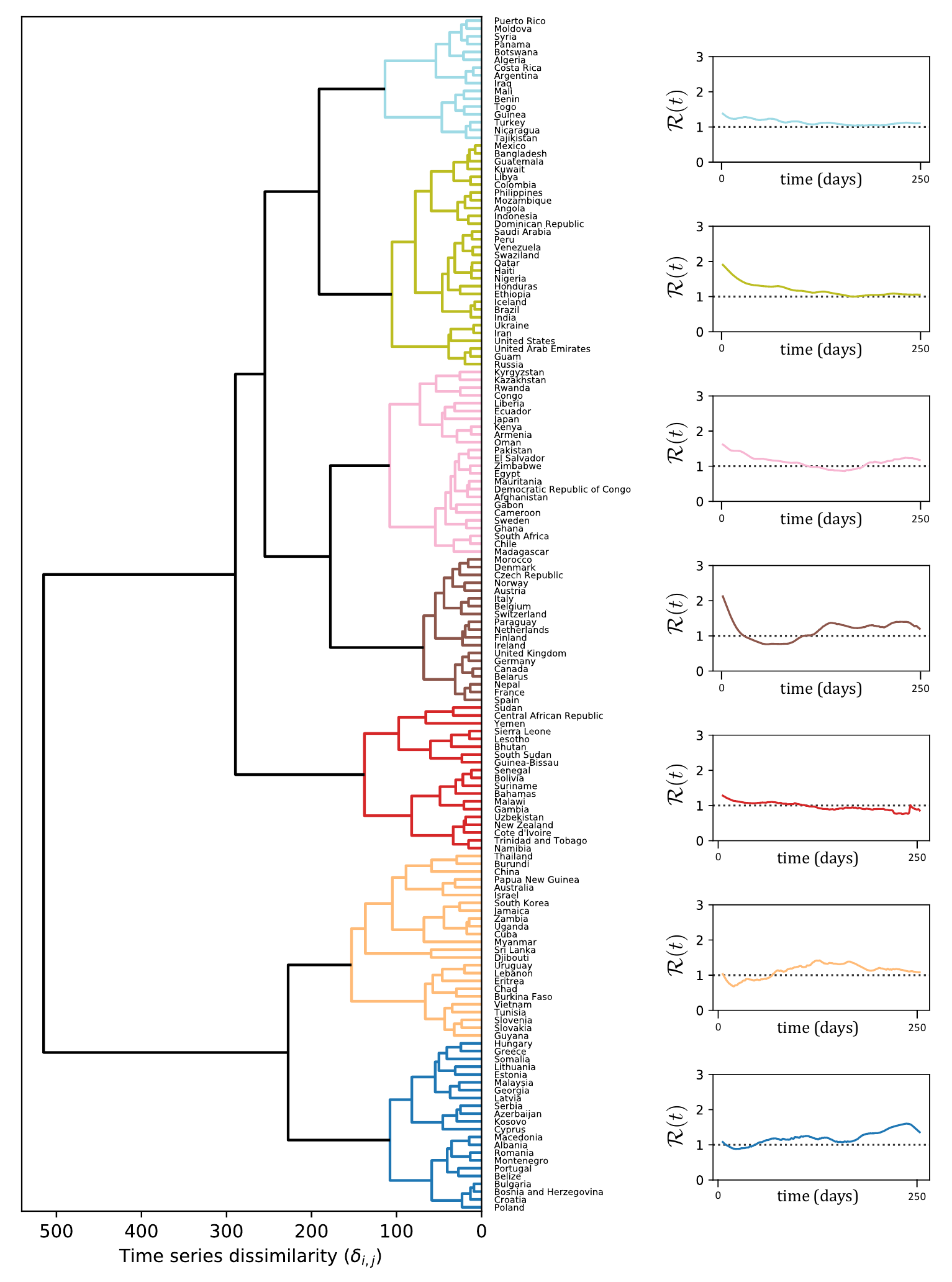}
    \caption{\footnotesize The left figure depicts the dendrogram obtained from the DTW dissimilarity between the $\mathcal{R}(t)$ time series. It is possible to obtain a certain number of clusters by putting a threshold at a specific dissimilarity value. In the example, we choose the threshold equal to 130. The right figures depict the average $\mathcal{R}(t)$ of countries in the same cluster.}
    \label{fig:dendrogram}
\end{figure}

\begin{figure}[t!]
	\centering
  \includegraphics[width=14 cm]{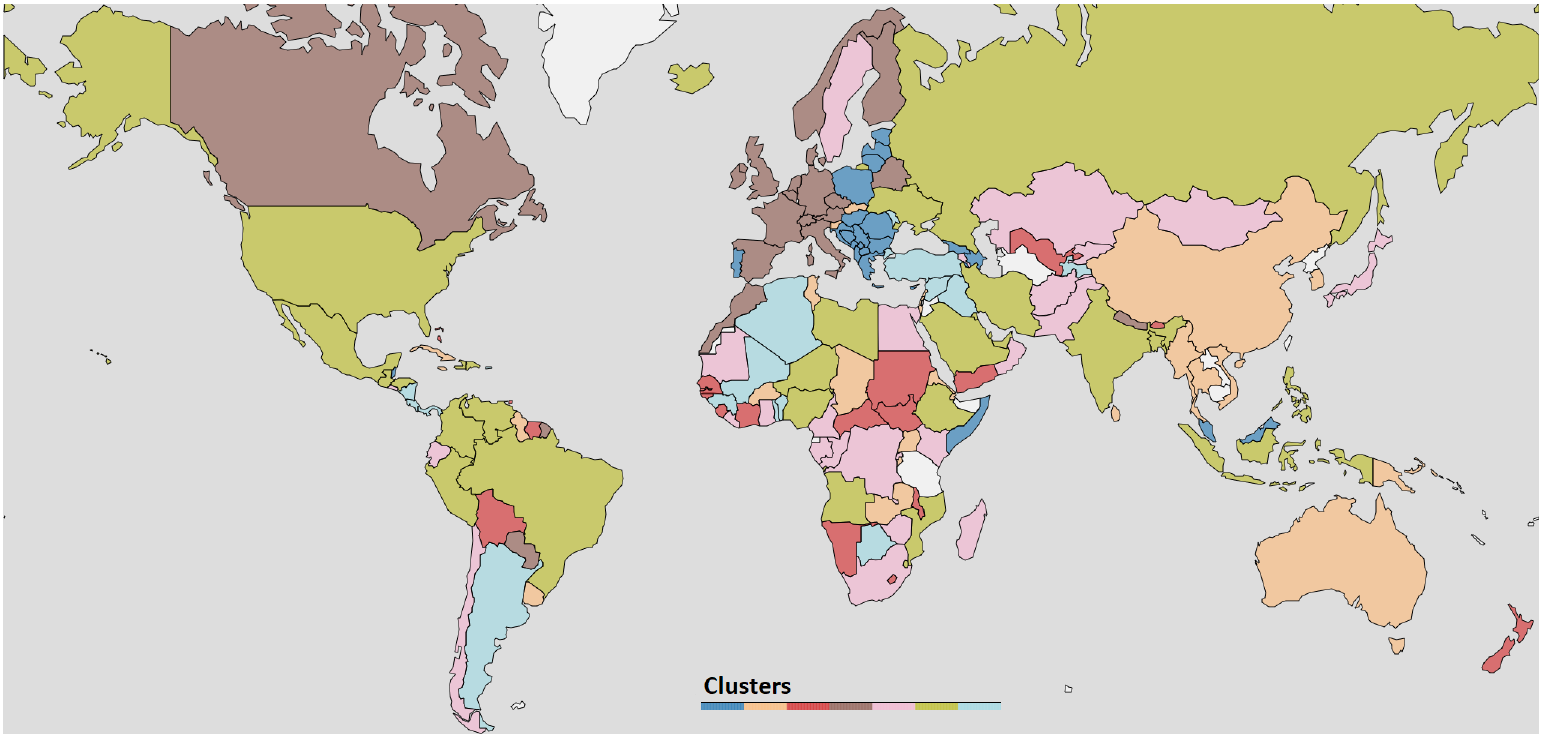}
    \caption{\footnotesize Visualization on the World map of the clusters obtained from the dissimilarity of the $\mathcal{R}(t)$ curves.}
	\label{fig:world}
\end{figure}

The dendrogram to the left of Figure~\ref{fig:dendrogram} depicts the result of the clustering procedure, based on the DTW dissimilarities between the $\mathcal{R}(t)$-curves estimated according to Eq.~(\ref{9}).
In particular, the dendrogram illustrates how two leaves $i$, $j$ (\textit{i.e.}, the $\mathcal{R}(t)$-curves of countries $i$ and $j$) are merged together as soon as the threshold $\delta_\text{max}$ becomes larger than their DTW dissimilarity value $\delta_{i,j}$.
There is no unique way of selecting an optimal $\delta_\text{max}$, but it rather depends on what level of resolution of the clustering partition is amenable for a meaningful exploration of the structure underlying our data.
In our case, we selected a $\delta_\text{max}=150$ that gave rise to seven clusters, depicted in different colors in Figure~\ref{fig:dendrogram}.
On the right hand side of Figure~\ref{fig:dendrogram}, we report $\mathcal{R}(t)$ averaged over all the countries in the same cluster.
To facilitate the interpretation of the results, in Figure~\ref{fig:world} we depict the same clustering partition obtained for $\delta_\text{max}=150$ on the political world map. 

The 1\textsuperscript{st} cluster (light blue) contains countries mostly from Africa, South America and Middle East. 
The average $\mathcal{R}$ curve of the countries in the light blue cluster (top-right of Figure~\ref{fig:dendrogram}) shows that the reproduction number is always very low, but consistently above one.
A possible explanation is that in those countries communities are more isolated and there are less travels and exchanges between them, making the infection to spread slower.

In 2$\textsuperscript{nd}$ cluster (green) the average $\mathcal{R}$ curve also stays always above one, but it starts from a higher value $\mathcal{R}_0$. It is important to notice that this cluster includes large countries, like India, Brazil, United States and Russia.
In these countries, the time series of new cases $X(t)$ have a particular profile since they are a combination from widely separated areas where the infection outbreak followed  different courses. For instance, in the U.S., the waves in New York and California are almost in opposite phase.

The 3$\textsuperscript{rd}$ cluster (pink) contains countries where the first wave is very long and it took a considerable amount of time to bring the $\mathcal{R}$ curve below 1. A second wave is slowly emerging in the Northern autumn. An atypical member of this cluster is Sweden, which experienced a second wave in the summer that appeared almost as a continuation of the first wave, and then a strong third wave in the fall that is synchronous with the second wave for the rest of West-Europe (cluster 4).

The 4$\textsuperscript{th}$ cluster (brown) mostly contains West-European countries, characterized by a strong first wave that was brought down quickly and a second wave that begun in the fall.
The average $\mathcal{R}$ curve is characterized by strong variability: it starts from a very high value and goes quickly below 1, to raise again quickly in the summer.

Similarly to the 1\textsuperscript{st} cluster, the 5$\textsuperscript{th}$ cluster (red) contains South American, African countries, and New Zealand. However, a key difference from 1$\textsuperscript{st}$ cluster is that in this case the $\mathcal{R}$ curve goes and remains below 1 during the Northern fall and autumn.

Finally, clusters 6 and 7 differ from the others by exhibiting initial $\mathcal{R}$  close to, or even lower than, one. For some countries in Cluster 6 this is an artifact of the averaging over all the countries in the cluster which includes some countries like China, Australia, and South Korea, which started out with quite high ${\cal R}$, but brought it down very rapidly through strong interventions  \cite{Rahman2020}.  
The common characteristic feature for the cluster is an ${\cal R}(t)$ above 1 during the Northern summer, but a reduction in the fall, which is the opposite of what has been observed in Western Europe and Canada.  
Cluster 7, on the other hand, contains most East-European countries, where the reproduction number was very low during the spring, but increased rapidly  after the summer.

\subsection{Exploring the parameter space of the  oscillator model}
\label{sec:params_exploration}
%As demonstrated in the previous subsection, 
The data for the incidence rate $X(t)$ in the world's countries show a wavy pattern consisting of one to three maxima during the first year of the pandemic evolution. However, the duration, relative strength, and separation between the waves vary substantially among countries and regions of the world. The total cumulative number of confirmed cases and deaths per million inhabitants can also vary by an order of magnitude or more among countries comparable to economic development, culture, and the healthcare system. Rypdal and Rypdal \cite{RR2020} demonstrated this for the first wave of the pandemic in a sample of 73 countries and discussed the significant differences in death toll between the two neighboring countries, Sweden and  Norway. At the time of writing this paper, we are four months further into the pandemic. The picture has changed dramatically, with secondary and tertiary waves developing in many countries.
%, as shown in the preceding subsection.

\begin{figure}[t!]
\centering
\includegraphics[width=14 cm]{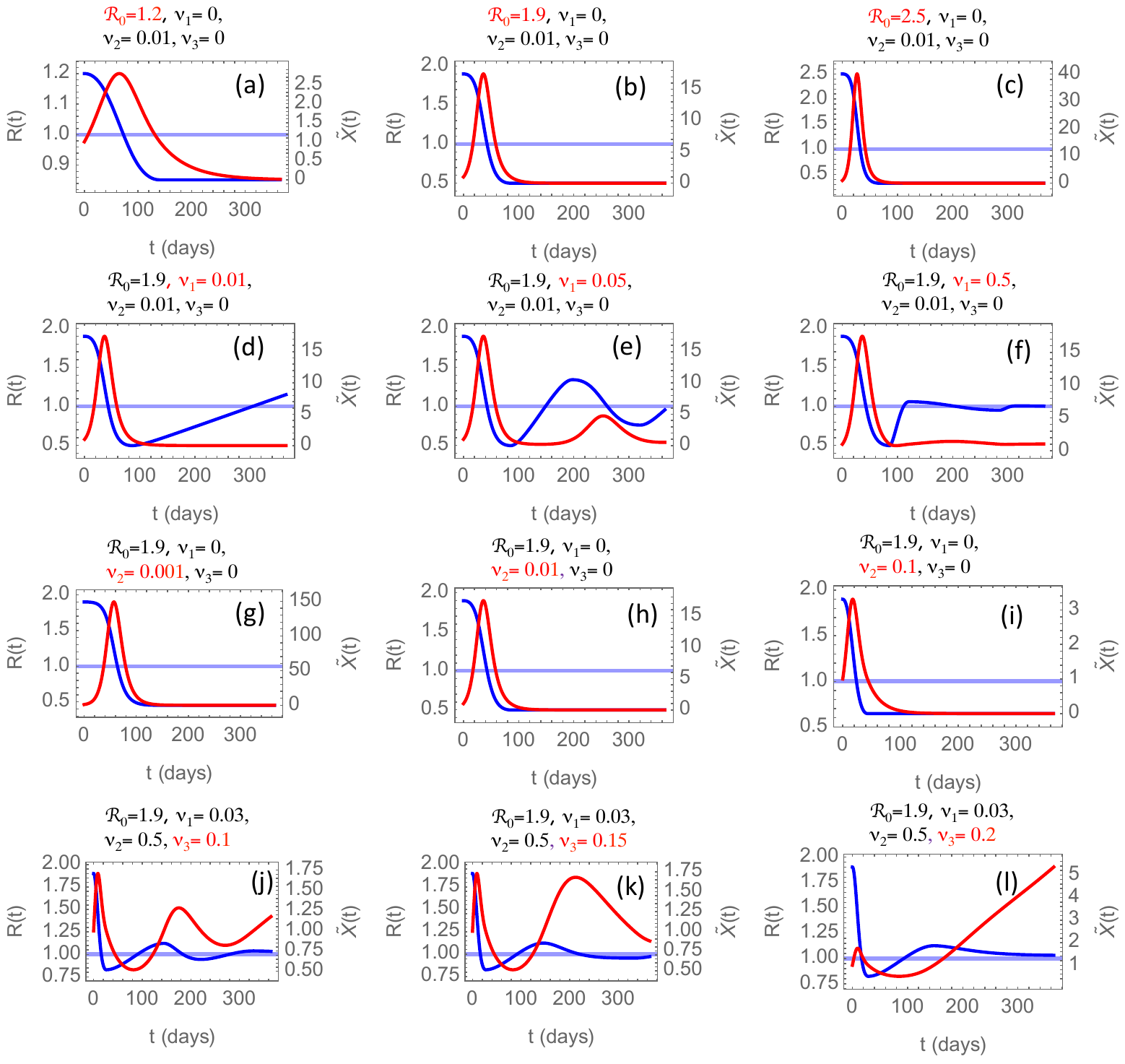}
\caption{Blue curves show the evolution of ${\cal R}(t)$ and red curves $\tilde{X}(t)$ as solutions of Eqs.(\ref{19}) and (\ref{20}) with $\tilde{X}={\cal R}\tilde{I}$ and $\nu$ given by Eq.\;(\ref{27}), initial conditions $\tilde{X}(0)=1$, ${\cal R}(0)={\cal R}_0$, and parameters ${\cal R}_0,\, \nu_1,\, \nu_2,\, \nu_3$ as indicated in the figures.}\label{fig:RXplotarray}
\end{figure}  

In Figure~\ref{fig:RXplotarray}, we have summarized some of the conclusions drawn from numerical solutions of the model proposed in Section~\ref{sec:closed_model}, obtained by varying the model parameters. In all simulations, we have chosen the time origin $t=0$ to be the first time the incidence rate $X(0)$ crosses the threshold value $X^*$. Hence, $\tilde{X}(0)=1$ for all simulations. The incidence rate measured on the right-hand axis in the figures is measured in units of the threshold $X^*$.

\subsubsection{The effect of the initial reproduction number}\label{sec:effect of R0}
In the first row of panels, Figure~\ref{fig:RXplotarray}(a)--(c), we consider the effect of changing the initial reproduction number ${\cal R}_0$. From the reconstructed ${\cal R}(t)$-curves, we observe that ${\cal R}_0$ varies considerably among countries and regions. Low values just above ${\cal R}_0=1$ are common in developing countries in South America, sub-Sahara Africa, and India. Several factors may contribute to this; lower mobility of people, a younger population, and a warmer climate. For these countries, we typically observe a  slower rise and decay of the first wave, and the wave is generally weaker than in industrialized countries where ${\cal R}_0$ varies in the range 2.0--2.5. In Figure~\ref{fig:RXplotarray}(a)--(c), we have changed ${\cal R}_0$, keeping $\nu_1$, $\nu_2$, $\nu_3$ constant. By choosing $\nu_1=0$, and $\nu_3=0$ we consider countries that respond slowly to an incidence rate below the threshold and show little fatigue, which may be characteristic for developing countries for which Figure~\ref{fig:RXplotarray}(a) may be relevant. 
In these panels, the choice $\nu_2=0.01$ is somewhat arbitrary but yields a rather stretched-out and low-amplitude first wave typical for those countries. For higher ${\cal R}_0$, the first wave is higher in amplitude and shorter, like what we have seen in China. Here $\nu_1=\nu_3=0$ signify that the relaxation rate and fatigue have been sufficiently low to prevent ${\cal R}$ from increasing after it has stabilized below 1. The maximum incidence rate $\tilde{X}=X/X^*$ in panels (b) and (c) is high; in the range 15-40. In panel (i), where the parameters are the same as in (b) except for $\nu_2=0.1$ being ten times higher, shows $\tilde{X}\approx 3$, which, as we will see later, is representative for China.

\subsubsection{The effect of the relaxation rate}
In the second row, we vary the relaxation rate $\nu_1$ while keeping the strike-down rate fixed at $\nu_2=0.01$. The result is that as $\tilde{X}$ drops below the threshold after about 45 days, $\mathcal{R}(t)$ starts to rise and grow well beyond 1. How fast this happens, depends on $\nu_1$. In the phase when ${\cal R}>1$, $\tilde{X}(t)$ will also start growing, and when it crosses the threshold $\tilde{X}=1$, the strike-down sets in again, and we enter a new cycle. With $\nu_1=0.01$ the first cycle takes almost 500 days, while it takes considerably less time in most countries, suggesting a higher $\nu_1$. In panel (e) we increase $\nu_1$ by a factor 5 and observe then two cycles within the first year, and in panel (f) another increment by a factor 10 almost eliminates the next waves. 
This faster relaxation to the equilibrium ${\cal R}=\tilde{X}=1$ when the relaxation rate is high may appear counter-intuitive. After all, it leads to a rapid increase of ${\cal R}$ once $\tilde{X}$ has dropped below the threshold. However, the faster rise of ${\cal R}$ also leads to a faster rise of ${\cal X}$ beyond the threshold and to a faster strike-down of ${\cal R}$ back towards 1, \textit{i.e.}, to faster damping of the oscillation. This observation suggests that the strong second wave of the epidemic evolving in Europe in the fall of 2020 is not caused by the relaxation of social interventions during the summer but is  caused by something else.

\subsubsection{The effect of the intervention rate}
 A suspected candidate could be the intervention rate $\nu_2$, which is varied in the third row, panels (g)--(i). However, we observe that the main effect of increasing $\nu_2$ is to decrease the amplitude of the oscillation in $\tilde{X}$ in inverse proportion to $\nu_2$. In this row, we have kept $\nu_1=0$, resulting in relaxation to a time-asymptotic ($t\rightarrow \infty$) equilibrium ${\cal R}_{\infty}<1$, $\tilde{X}_{\infty}=0$. This is in contrast to the second row ($\nu_1>0$), where this equilibrium is ${\cal R}_{\infty}=1$, $\tilde{X}_{\infty}=1$. These two equilibria correspond to fundamentally different strategies to combat the epidemic. The one without the relaxation mechanism ($\nu_1=0$) corresponds to the strike-down strategy, where the goal is to eliminate the pathogen without obtaining herd immunity in the population. The one with $\nu_1>0$, allowing relaxation of interventions when the incidence rate dips below the threshold, will end up with a constant incidence rate at the threshold value and thus a linearly increasing cumulative number of infected until this growth is non-linearly saturated by herd immunity.

 \subsubsection{The effect of the fatigue  rate}
 The effect of a non-zero fatigue rate is to bring the effective strike-down rate to zero as $t\rightarrow \infty$. The  solution of the system Eqs.\;(\ref{19})--(\ref{20}) as $t\rightarrow \infty$ is that ${\cal R}\rightarrow 1+\nu_3$ and $\tilde{X}\approx \exp{(\nu_3 t)}$. Of course, this blow-up is prevented by herd immunity, which will reduce the effective ${\cal R}$ to zero when most of the population has been infected. The effect of increasing immunity in the population is not included in Eq.\;(\ref{17}), and hence the model makes sense only as long as the majority of the population is still susceptible to the disease.
 Nevertheless, the last row in Figure \ref{fig:RXplotarray} shows that increasing intervention fatigue represented by non-zero $\nu_3$ may increase the second and later waves' amplitude and duration. 
 For sufficiently large $\nu_3$ the second wave's amplitude and duration can become greater than the first. The situations shown in panels (k) and (l) are observed in European countries and are caused by $\nu_3\approx 1$. One partial explanation of the second wave's higher amplitude than the first, as observed in many countries, is a considerably higher testing rate. The testing rate, however, cannot explain the considerably longer duration of the second wave. This prolonged duration shows up both in the observed data and in this model, when the fatigue rate is increased.

\subsection{Results from fitting the oscillator model to incidence data in different countries}
\label{sec:fit}

\begin{figure}[t!]
	\centering
  \includegraphics[width=14 cm]{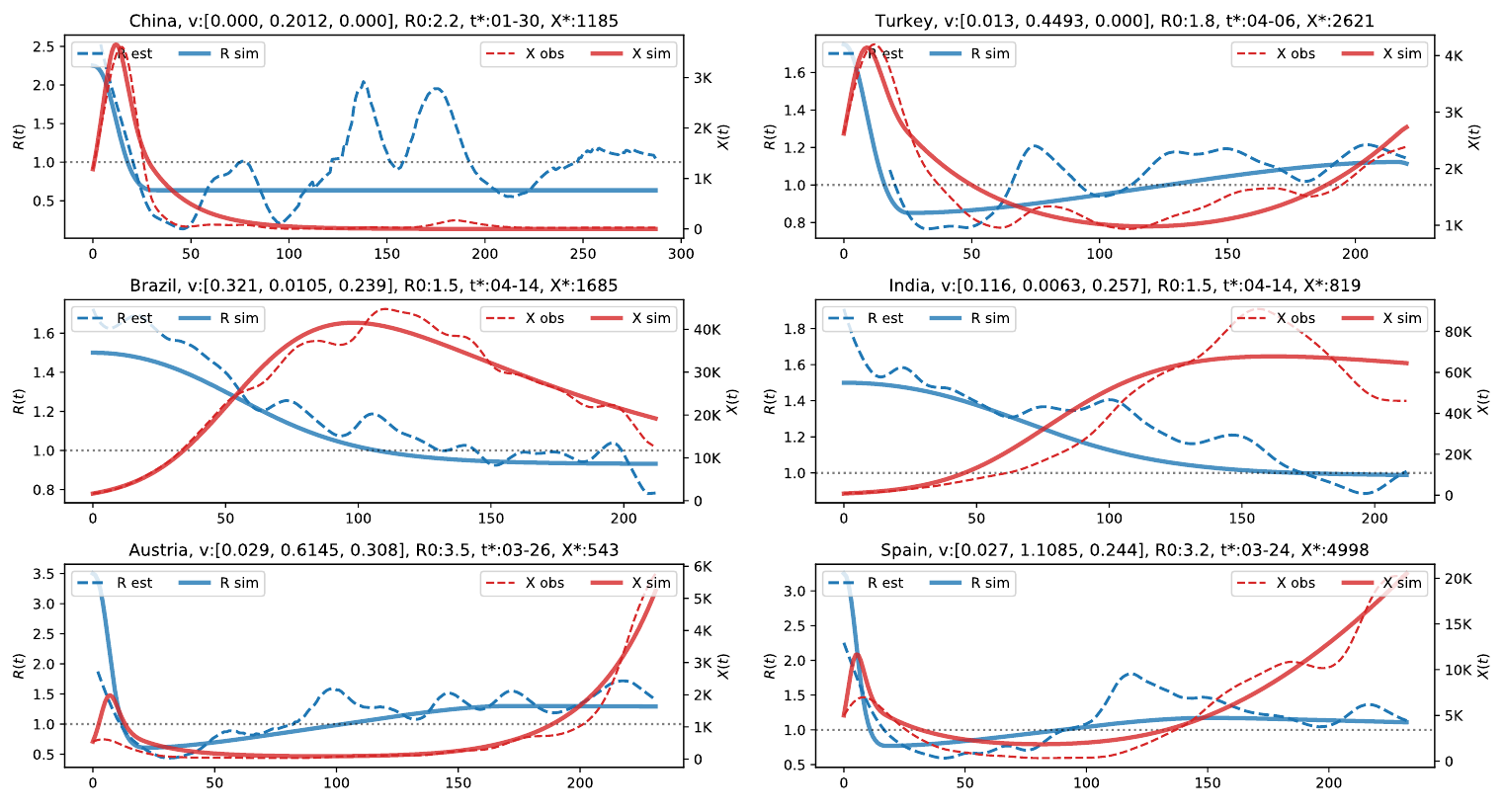}
    \caption{\footnotesize Comparison between observed daily cases $X(t)$ (red dashed lines), $X(t)$ simulated from model (solid red lines), $\mathcal{R}(t)$ estimated from the data using the inverted SIR model (dashed blue line), $\mathcal{R}(t)$ simulated from the model.}
	\label{fig:Model_estimate_comp}
\end{figure}

Figure~\ref{fig:Model_estimate_comp} depicts for selected countries the reported daily new cases (dashed red line), the daily new cases simulated by the oscillator model (solid red line), the $\mathcal{R}(t)$ curve estimated using Eq.\;(\ref{9}) (dashed blue line), and the $\mathcal{R}(t)$ curve simulated by the proposed close model (solid blue line).
On the top of each graph, we report for each country the fitted values of $[\nu_1$, $\nu_2$, $\nu_3]$, the initial $\mathcal{R}_0$, and the date $t^{*}$ that identifies $X^{*} = X(t^{*})$.
On the horizontal axis, 0 corresponds to $t^{*}$, the left vertical axis indicates the value of the reproduction number, the right vertical axis indicates the number of new daily cases. 

The first row in the array of panels shows results for China and Turkey. The parameters estimated for China is comparable to those in Figure~\ref{fig:RXplotarray}(i). The peak incidence rate $X_{\text{max}}\approx 3,500$  for China is about 3 times the threshold incidence $X^*=1,185$, similar to what is observed in Figure~\ref{fig:RXplotarray}(i). For Turkey, the evolution of ${\cal R}(t)$ is initially rather similar to that of China, and the shape of the $X(t)$-curve is also rather similar. But while the model-fitted ${\cal R}(t)$ converges to a fixed value ${\cal R}_{\infty}<1$, and $X(t)$ to 0 after a few months in China, ${\cal R}(t)$ in Turkey grows slowly greater than 1, and a second wave in $X(t)$ develops. This wave has an amplitude approximately the same as the first, but lasts longer (not shown in the figure), similar to what is shown in Figure~\ref{fig:RXplotarray}(k). This rise is the result of fundamental differences in the estimated model parameters: $\nu_1$ is zero for China but non-zero for Turkey.
We also notice that China and Turkey belong to different clusters in the dendrogram in Figure~\ref{fig:dendrogram}. 
The second wave for Turkey is not created by a finite fatigue rate, since $\nu_3=0$ both for Turkey and China; it is created by a finite relaxation rate $\nu_1$. Importantly, this relaxation rate cannot create a second wave that is stronger than the first, it only gives rise to a damped oscillation that ends up in the equilibrium ${\cal R}=1$, $X=X^*$.

The second row shows two countries, Brazil and India belonging to the second cluster, represented in green in Figure~\ref{fig:dendrogram}. The initial reproduction number is low, ${\cal R}_0=1.5$ for both countries, and the strike-down parameter is also low, $\nu_2\approx 0.01$, leading to a strong and long first wave, which is not yet completely over in November 2020. The fatigue rate of $\nu_3\approx 0.2$ also contributes to increasing the amplitude and the long-lasting downward slope of the first wave. 

In the third row, we show the typical pattern for Europe, the fourth cluster (brown) in the dendrogram, with Austria and Spain as examples. There is a rather short first wave accompanied by a rapid drop in ${\cal R}(t)$ due to the almost universal lockdown in March 2020. There is a rather slow relaxation of the interventions throughout the summer, finally leading to ${\cal R}$ stabilizing in the range 1.2-1.5. The inevitable result is the rise of a second wave, growing stronger and longer than the first, as shown in Figure~\ref{fig:RXplotarray}(k) and (l). At the time of writing,  interventions again have started to inhibit the growth, but they are weaker than in the spring, as reflected by the fatigue rates in the range $\nu_3\sim 0.2-0.3$. 
Indeed, the predictions of the oscillator model with the estimated rates is that the second wave will blow up in the spring of 2021 to levels where herd immunity will limit the growth. This is before vaccines are likely to play an important r\^{o}le, so a more probable scenario is that governments will  reverse the fatigue trend and invalidate the model as a prediction for the future. Tendencies in this direction is observed in Europe at the time of writing.

\section{Discussion and conclusions}\label{sec:discuss}
The geographic distribution of countries belonging to different clusters shown in the map in Figure~\ref{fig:world},  and the associated  averaged ${\cal R}(t)$-curves in Figure~\ref{fig:dendrogram}, may serve as a crude road map to the global evolution of the pandemic throughout the spring and fall of 2020. One striking feature is some geographic clustering, which is most pronounced in Western Europe (brown) and  Eastern Europe (blue). A similar clustering is seen in the U.S. and Equatorial Latin America (green). In this paper, we have a focus on the strength, timing and duration of the second epidemic wave, and for this purpose the dendrogram helps us to identify those regions where there has been a pronounced second wave so far in the pandemic. These are those countries that belong to clusters exhibiting a period of ${\cal R}(t)<1$ in between periods of ${\cal R}>1$. From the  ${\cal R}(t)$-profiles in Figure~\ref{fig:dendrogram} those countries with the most pronounced second wave are Cluster 4 (brown) and 7 (dark blue), Western and Eastern Europe, respectively. The rise of the second wave here is due to the persistently high values of ${\cal R}$  during the period July - November. What distinguishes the two clusters is the course of the first wave. In Western Europe there was a strong first wave associated with  high ${\cal R}$, and it affected strongly older age groups which resulted in high case fatality ratio (CFR). The second wave has affected all ages and so far the death numbers have been much lower than in the first. In Eastern Europe the first wave was very weak, but the second has been strong and with considerably higher CFR than in the countries further West. 

The main result in this paper is to demonstrate that the varying courses of the epidemic depicted via the seven characteristic ${\cal R}(t)$ curves shown in Figure~\ref{fig:dendrogram} to some extent ca be understood in terms of the interplay between three social responses to the epidemic activity; the relaxation of interventions when the activity is low, the intensification of interventions when activity becomes high, and the intervention fatigue which develops with time. Figures~\ref{fig:RXplotarray} and ~\ref{fig:Model_estimate_comp}  suggest that most country-specific epidemic curves can be qualitatively reproduced by a simple mathematical model involving these three responses. The value of this insight is that, in spite of the immense complexity and diversity of the dynamical response triggered by this new pathogen, there are some universal governing principles that will determine the final outcome in the years to come. 

The model devised here could of course be run to make projections further ahead than one year from the onset of the epidemic, as done in Figure ~\ref{fig:RXplotarray}. It would show a blow-up of all solutions for which the fatigue parameter is non-zero, and would be unrealistic for several reasons. One is that the linearity approximation would break down as herd immunity will start to bring the effective reproduction number down. Another is that the intervention fatigue model most likely will fail when the epidemic activity becomes sufficiently high. We have already have seen signs in this direction in many European countries where partial lockdowns and mass testing again have succeeded in ``bending the curve'' to an extent that is not described by the model. Finally, mass-vaccination will hopefully become a  real game-changer in the year to come.

\end{document}